\documentclass{prop2015}
\usepackage[english]{babel}
\keywords{phase transition, disorder, critical behavior}
\title{Emerging critical behavior at a first-order phase transition rounded by disorder}
\author[A.\,K. Ibrahim]{Ahmed K. Ibrahim\inst{1}}
\author[T. Vojta]{Thomas Vojta\inst{1}\footnote{Corresponding author\quad E-mail:~\textsf{vojtat@mst.edu}}}
\address[1]{Department of Physics, Missouri University of Science and Technology, Rolla, MO 65401, USA}
\begin{abstract}
We investigate the two-dimensional four-color Ashkin-Teller model by means of large-scale Monte-Carlo simulations.
We demonstrate that the first-order phase transition of the clean system is destroyed by random disorder introduced
via site dilution. The critical behavior of the emerging continuous transition
belongs to the clean two-dimensional Ising universality class, apart from logarithmic
corrections. These results confirm perturbative renormalization-group predictions; they also agree with recent
findings for the three-color case, indicating that the critical behavior is universal.
\end{abstract}
\shortabstract
\begin{document}
\maketitle
\section{Introduction}

At a first-order phase transition, two distinct thermodynamic phases have the same free energy density
and thus coexist macroscopically. Is such phase coexistence still possible if the system contains
random disorder that locally favors one phase over the other (so-called random-$T_c$ or random-mass
disorder)? Building on earlier work for random fields \cite{ImryMa75}, Imry and Wortis
\cite{ImryWortis79} compared the possible free-energy gain from forming a domain that takes advantage 
of the disorder with the free-energy cost
of its domain wall. They found that the formation of finite-size domains is favored in dimensions
$d\le 2$, even for arbitrarily weak disorder. (If the randomness breaks a continuous symmetry,
the marginal dimension is $d = 4$.)
 This destroys the macroscopic phase coexistence
and with it the first-order phase transition. After further work \cite{HuiBerker89},
this result was rigorously proven by Aizenman and Wehr \cite{AizenmanWehr89}.

What happens to a system whose first-order phase transition is unstable against disorder?
Is there an intermediate phase? Is the transition completely destroyed by smearing;
or does it become continuous? If the latter, in what universality class is the emerging
critical point?

The effects of disorder on first-order phase transitions have received less attention than the
corresponding effects on continuous transitions (see, e.g., Refs.\ \cite{Vojta06,Vojta10} for reviews).
As a result, the above questions are still being debated even for simple model systems such as
the two-dimensional Ashkin-Teller model. The classical $N$-color Ashkin-Teller model
\cite{AshkinTeller43,GrestWidom81,Fradkin84,Shankar85} is made up of $N$ Ising models
that are coupled via four-spin interactions. For more than two colors ($N>2$), the clean
Ashkin-Teller model undergoes a first-order phase transition from a magnetically ordered 
phase at low temperatures to a paramagnetic phase at high temperatures.

In two dimensions, the
first-order transition cannot survive in the presence of disorder; the two-dimensional
$N$-color Ashkin-Teller model is thus a prototypical system for studying the questions posed above.
Perturbative renormalization group calculations \cite{Murthy87,Cardy96,Cardy99} predict
that disorder rounds the first-order transition of the clean-Ashkin-Teller model to a continuous
one which, somewhat surprisingly, is in the \emph{clean} Ising universality class with additional
logarithmic corrections to scaling. Early numerical simulations \cite{BellafardKatzgraberTroyerChakravarty12,BellafardChakravartyTroyerKatzgraber15}
of small systems reported nonuniversal critical behavior. However, a recent high-accuracy study
\cite{ZWNHV15} of the disordered three-color Ashkin-Teller model provided strong evidence in favor
of the scenario predicted by the perturbative renormalization group \cite{Murthy87,Cardy96,Cardy99}.

In the present paper, we test the universality of the emerging critical behavior in the
disordered Ashkin-Teller model by studying the four-color case and comparing it to the existing
simulations for three colors as well as the renormalization group findings. Specifically, we report results of large-scale
Monte Carlo simulations of the site-diluted two-dimensional four-color Ashkin-Teller model with up
to $2240^2$ sites. The paper is organized as follows. In Sec.\ \ref{sec:model}, we introduce
the model and summarize the predictions of the perturbative renormalization group.
Section \ref{sec:MC} is devoted to the Monte-Carlo simulation results. We summarize and
conclude in Sec.\ \ref{sec:conclusions}.

\section{Diluted Ashkin-Teller model}
\label{sec:model}

The two-dimensional $N$-color Ashkin-Teller model \cite{GrestWidom81,Fradkin84,Shankar85}
is defined on a square lattice of $L^2$ sites. Each lattice site $i$ contains $N$ Ising spins $S_i^\alpha=\pm 1$,
distinguished by the ``color''-index $\alpha=1\ldots N$. In the absence of disorder, the
Hamiltonian reads
\begin{equation}
H = -J \sum_{\alpha=1}^N \sum_{\langle ij \rangle} S_i^{\alpha} S_j^{\alpha}
    -\epsilon J \sum_{\alpha<\beta} \sum_{\langle ij \rangle} S_i^{\alpha} S_j^{\alpha} S_i^{\beta} S_j^{\beta}~.
\label{eq:HAT}
\end{equation}
It can be understood as $N$ identical Ising models that are coupled via their energy
densities. $\langle ij \rangle$ denotes the sum over pairs of nearest neighbor sites on the lattice;
$J>0$ is a ferromagnetic interaction; and $\epsilon\ge 0$ parameterizes the strength
of the inter-color coupling.
The clean two-color model (the original model proposed by Ashkin and Teller \cite{AshkinTeller43})
undergoes a continuous phase transition from a magnetically ordered (Baxter) phase at
low temperatures to a paramagnetic phase at high temperatures. In the Baxter phase, the spins of each color
order ferromagnetically  w.r.t. each other but the relative orientation of the colors is arbitrary (see, e.g., Ref.\ \cite{Baxter_book82}).
The critical behavior of the two-color model is nonuniversal, i.e., the critical exponents change continuously with $\epsilon$.
We are interested in the case of three or more colors for which the phase transition between the paramagnetic and Baxter phases
is of first order \cite{GrestWidom81,Fradkin84,Shankar85}.

We introduce quenched disorder into the Hamiltonian (\ref{eq:HAT}) by means of site dilution. This means,
a fraction $p$ of the lattice sites are randomly replaced by vacancies on which the spins $S_i^\alpha$ for
all colors are removed. Site dilution is a microscopic realization of random-$T_c$ or random mass disorder,
i.e., disorder that locally favors one phase over the other but does not break any of the spin symmetries.

Murthy \cite{Murthy87} and Cardy \cite{Cardy96,Cardy99} studied the phase transition of the
$N$-color Ashkin-Teller model with random-$T_c$ disorder by means of a perturbative renormalization group.
They found that the renormalization group flow on the critical $(\epsilon,\Delta)$-surface asymptotically
approaches the clean Ising fixed point $\epsilon=0,\Delta=0$. (Here, $\Delta$ is a measure of the disorder
strength.) This implies a continuous transition that is, surprisingly, in the clean 2D Ising universality
class, apart from logarithmic corrections analogous to those occurring in the disordered Ising model
\cite{DotsenkoDotsenko83,Shalaev84,Shankar87,Ludwig88}. The following finite-size scaling behavior
has been derived \cite{MazzeoKuhn99,HTPV08,KennaRuizLorenzo08}.
At criticality ($T=T_c$), the specific heat shows a characteristic double-logarithmic dependence
on the system size,
\begin{equation}
C \sim \ln \ln L~.
\label{eq:C_lnln}
\end{equation}
Magnetization and susceptibility at $T_c$ (averaged over all $N$ colors) behave as
\begin{eqnarray}
M  &\sim& L^{-\beta/\nu}\, (1+b_M/\ln L)~,
\label{eq:M_ln}\\
\chi &\sim& L^{\gamma/\nu}\,(1+b_\chi/\ln L)~,
\label{eq:chi_ln}
\end{eqnarray}
where $\gamma/\nu=7/4$ and $\beta/\nu=1/8$ as in the clean Ising model, and $b_M$ and $b_\chi$ are constant.
Any quantity $R$ of scale dimension zero behaves as
\begin{eqnarray}
R &=& R^\ast \,(1+b_R/\ln L)~,
\label{eq:R_ln}\\
dR/dT  &\sim& L^{1/\nu}(\ln L)^{-1/2}\, [1+O(1/(\ln L))]
\label{eq:dRdT_ln}
\end{eqnarray}
with the clean Ising exponent $\nu=1$. The finite-size scaling forms (\ref{eq:R_ln}) and (\ref{eq:dRdT_ln})
hold for the Binder cumulants
\begin{equation}
g_\textrm{av} = \left[1 - \frac {\langle m^4\rangle}{3\langle m^2\rangle^2} \right]_\textrm{dis}~, \quad
g_\textrm{gl} = 1 - \frac {[\langle m^4\rangle]_\textrm{dis}}{3[\langle m^2\rangle]^2_\textrm{dis}} ~.
\label{eq:Binder}
\end{equation}
Here, $\langle \ldots \rangle$ denotes the thermodynamic (Monte-Carlo) average while $[\ldots ]_\textrm{dis}$
stands for the disorder average. We need to distinguish average and ``global'' versions of this quantity, depending
on when the disorder average is performed.
The scaling forms (\ref{eq:R_ln}) and (\ref{eq:dRdT_ln})
also hold for the correlation length ratios $\xi_\textrm{av}/L$ and $\xi_\textrm{gl}/L$.
In our simulations, the correlation lengths are computed from the second moment of the spin correlation function
$G(\mathbf{r}) = (1/L^2) \sum_{i,j,\alpha} \langle S_i^\alpha S_j^\alpha \rangle \delta(\mathbf{r}-\mathbf{r}_{ij})$ \cite{CooperFreedmanPreston82,Kim93,CGGP01}.
They can be obtained efficiently from the Fourier transform $\tilde G(q)$ of the correlation function:
\begin{eqnarray}
\xi_\textrm{av} &=& \left[ \left(\frac{\tilde G(0) -\tilde G (q_\textrm{min})}{q_\textrm{min}^2 \tilde G(q_\textrm{min})} \right)^{1/2} \right]_\textrm{dis}~,
\label{eq:xi_av}\\
\xi_\textrm{gl} &=&  \left(\frac{[\tilde G(0) -\tilde G (q_\textrm{min})]_\textrm{dis}}{q_\textrm{min}^2 [\tilde G(q_\textrm{min})]_\textrm{dis}} \right)^{1/2}~.
\end{eqnarray}
Here, $q_\textrm{min}=2\pi/L$ is the minimum wave number that fits into a system of linear size $L$.

\section{Monte Carlo simulations}
\label{sec:MC}
\subsection{Method and overview}
\label{subsec:Method}

To simulate the thermodynamics of the four-color Ashkin-Teller model (\ref{eq:HAT}),
we employ an embedding algorithm analogous to that used in Refs.\ \cite{WisemanDomany95,ZWNHV15}.
It is based on a simple observation. If the spins of colors $\alpha=2,3$ and 4 are fixed,
the Hamiltonian (\ref{eq:HAT}) is equivalent to an Ising model for the spins $S_i^{(1)}$
with effective interactions $J_{ij}^\textrm{eff} = J+ \epsilon J ( S_i^{(2)}S_j^{(2)} + S_i^{(3)}S_j^{(3)} + S_i^{(4)}S_j^{(4)} )$.
This (embedded) Ising model can be simulated using any valid Monte-Carlo method.
Analogous embedded Ising models can be constructed for the spins $S_i^{(2)}$, $S_i^{(3)}$, and $S_i^{(4)}$.
By combining Monte-Carlo updates for all four embedded Ising models we obtain a valid
Monte-Carlo method for the entire Ashkin-Teller Hamiltonian.

Using this algorithm, we simulate systems with sizes from $35^2$ to $2240^2$ sites
with periodic boundary conditions and dilution $p=0.3$. All results are averaged over a large
number of disorder configurations (10,000 to 500,000), details will be given below.
We combine Wolff single-cluster updates \cite{Wolff89} with Swendsen-Wang multi-cluster
updates \cite{SwendsenWang87}. The latter help equilibrating small isolated clusters of sites
that occur for larger dilutions. Specifically, each full Monte Carlo sweep consists of
a Swendsen-Wang sweep for each color and a Wolff sweep (a number of single-cluster flips
such that the total number of flipped spins for each color equals the number of lattice sites).
We use 100 full Monte Carlo sweeps for equilibrating each sample (disorder configuration)
and 200 sweeps for measuring observables (one measurement per sweep).
The actual equilibration times are much shorter \cite{ZWNHV15}.
Biases in the observables due to the short measurement periods are overcome by using
improved estimators \cite{ZWNHV15}.

The Wolff and Swendsen-Wang algorithms are only valid as long as all effective interactions
of the embedded Ising models,
$J_{ij}^\textrm{eff} = J+ \epsilon J ( S_i^{(2)}S_j^{(2)} + S_i^{(3)}S_j^{(3)} + S_i^{(4)}S_j^{(4)} )$, 
 are not negative. In the worst case, the term in parenthesis can take the value $-3$. 
The coupling constant $\epsilon$ therefore must not exceed 1/3. In the production runs, we use the largest possible value,
$\epsilon=1/3$, because this leads to a strong first-order transition in the clean case.

\subsection{Results}
\label{subsec:Results}

To find the phase transition of the four-color Ashkin-Teller model with coupling $\epsilon=1/3$
and dilution $p=0.3$ we perform a series of simulation runs using linear system sizes $L=35$ to 2240.
The number of disorder realizations ranges from 500,000 for the smallest systems to 10,000 for the largest
ones. As usual, the critical temperature $T_c$ can be determined from the crossings of the Binder cumulant vs.
temperature curves for different system sizes (and analogously from the crossings of the reduced
correlation length curves $\xi/L$ vs. $T$).

\begin{figure}
\includegraphics[width=\columnwidth]{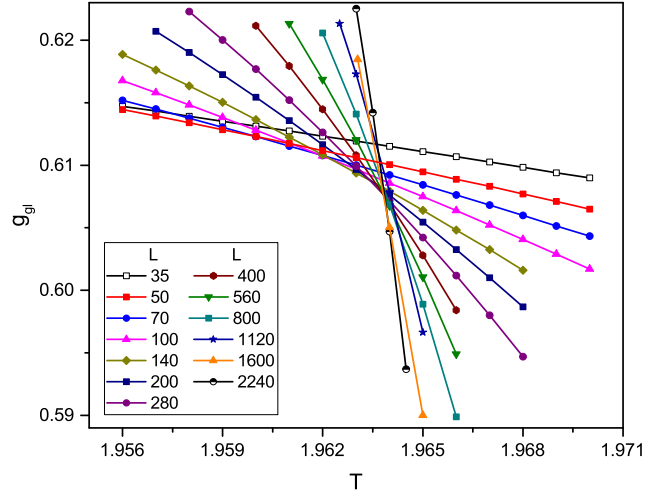}
\caption{Binder cumulant $g_\textrm{gl}$ vs.\ temperature $T$ of the site-diluted two-dimensional four-color Ashkin-Teller model with $p=0.3$ and $\epsilon=1/3$
        for different linear system sizes $L$.  The shift towards higher temperatures of the crossing point with increasing $L$
        is caused by corrections to scaling.}
\label{fig:bindercrossing}
\end{figure}

\begin{figure}
\includegraphics[width=\columnwidth]{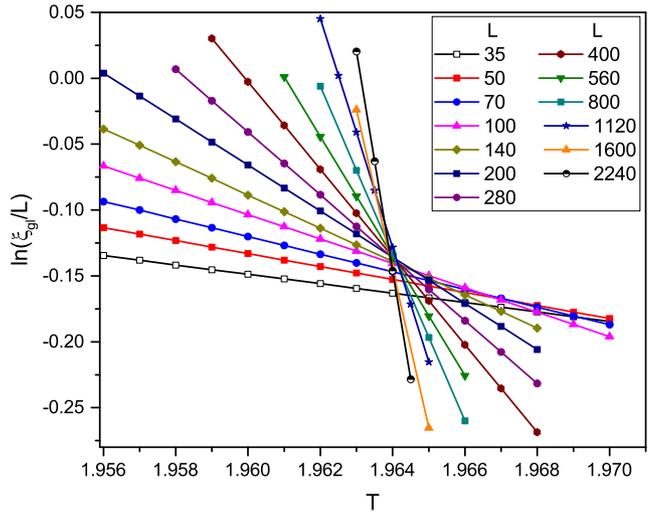}
\caption{Reduced correlation length $\xi_\textrm{gl}/L$ vs.\ temperature $T$
        for different linear system sizes $L$.  With increasing $L$, the crossings move to lower $T$, again
        indicating corrections to scaling.}
\label{fig:xicrossing}
\end{figure}

\begin{figure}
\includegraphics[width=\columnwidth]{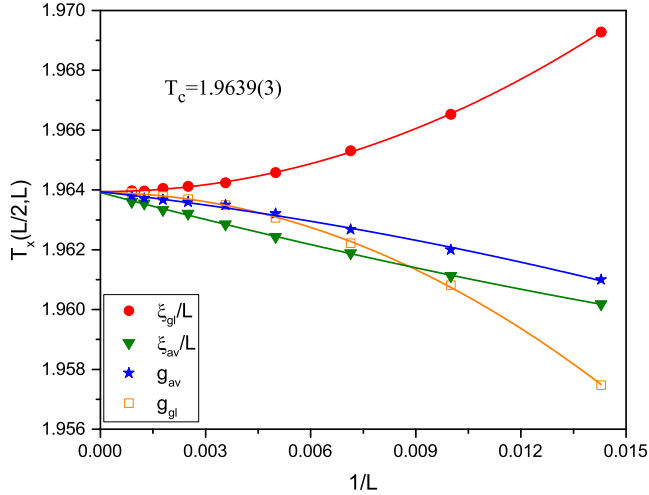}
\caption{Crossing temperatures $T_x(L/2,L)$ vs.\ inverse system size $1/L$. The solid lines are fits to
$T_x(L/2,L) =T_c + a L^{-b}$ yielding $T_c=1.9639(3)$. The error bars of $T_x$ are about a symbol size for the smallest $L$, they become
much smaller with increasing $L$.}
\label{fig:TxvsL}
\end{figure}

The Binder cumulant $g_\textrm{gl}$ as a function of temperature $T$ is shown in Fig.\ \ref{fig:bindercrossing},
and Fig.\ \ref{fig:xicrossing} presents the reduced correlation length $\xi_\textrm{gl}/L$ as a function of temperature $T$.
Similar graphs can be created for the quantities $g_\textrm{av}$ and $\xi_\textrm{av}/L$. In all cases, the crossing points
of their curves for different $L$ move with increasing $L$. This indicates that corrections to the leading scaling behavior
are important for the studied system sizes. To find the true, asymptotic value of the critical temperature,
we therefore need to extrapolate the crossing points to infinite system size. To do so, we find the crossing temperature $T_x(L/2,L)$
of the $g_\textrm{gl}$ vs.\ $T$ curves for system sizes $L/2$ and $L$ as well as the analogous crossing temperatures
for   $g_\textrm{av}$, $\xi_\textrm{gl}/L$, and $\xi_\textrm{av}/L$. The resulting dependence of the crossing temperatures on the system
size is presented in Fig.\ \ref{fig:TxvsL}.
The figure shows that all crossings approach the same temperature as $L$ increases. Fits to the heuristic relation
$T_x(L/2,L) =T_c + a L^{-b}$ yield a critical temperature of $T_c=1.9639(3)$ (the number in parentheses is an estimate of the error
of the last digits).

\begin{figure}
\includegraphics[width=\columnwidth]{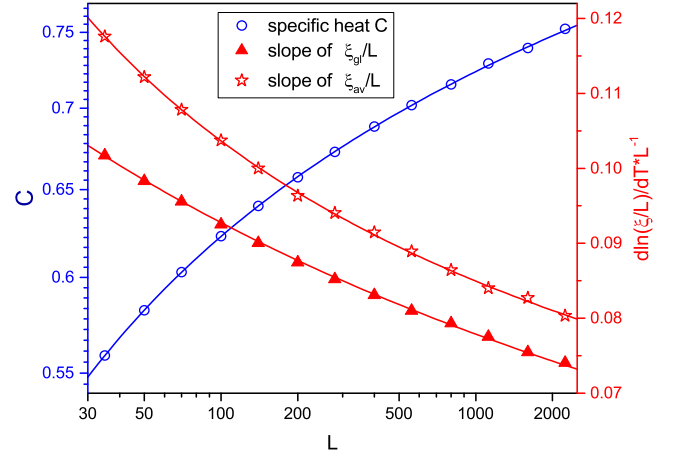}
\caption{Double-logarithmic plots of the specific heat $C$ as well as the slopes $L^{-1} d\ln(\xi_\textrm{gl}/L)/dT$
     and $L^{-1} d\ln(\xi_\textrm{av}/L)/dT$ vs.\
     system size $L$ at the critical temperature of $T_c=1.9639$. The error bars are smaller than the symbol sizes.
     The solid lines are fits to
     $a\ln[b\ln(cL)]$ for $C$ and $a[\ln(bL)]^{-1/2}$ for the slopes.}
\label{fig:CV}
\end{figure}

To determine the critical behavior, we now analyze the finite-size scaling properties of various observables right at $T_c$.
The data analysis follows Ref.\ \cite{ZWNHV15} and is based on the finite-size scaling forms (\ref{eq:C_lnln}) to (\ref{eq:dRdT_ln}).
Figure \ref{fig:CV} presents a double-logarithmic plot of specific heat $C$ vs.\ system size $L$.
The specific heat increases more slowly than a power law with $L$, as indicated by the downward curvature of the graph. In agreement with the
prediction (\ref{eq:C_lnln}), the data can be fitted well to the form $a \ln[b \ln(cL)]$ over the entire
size range. The fit is of high quality, giving a reduced error sum $\bar\chi^2 \approx 0.8$. (The reduced error sum of a fit of $n$ data points
($x_i,y_i$) to a function $f(x)$ having $q$ fit parameters is defined as $\bar\chi^2 = 1/(n-q)\sum_i [y_i-f(x_i)]^2/\sigma_i^2$
where $\sigma_i$ is the standard deviation of $y_i$.) For comparison, we have also attempted to fit the specific heat to the simple logarithmic
form $a\ln(bL)$ and to $C_\infty - a L^{-b}$. The latter function corresponds to power-law scaling with a negative finite-size scaling
exponent $\alpha/\nu=-b$. The power-law fit is of poor quality with a reduced error sum $\bar\chi^2 \approx 4.3$, and the simple logarithmic fit
is completely off, giving $\bar\chi^2 \approx 420$.

Figure\ \ref{fig:CV} also presents the slopes $d\ln(\xi_\textrm{gl}/L)/dT$ and
$d\ln(\xi_\textrm{av}/L)/dT$ of the normalized correlation lengths vs.\ temperature curves at $T_c$. We have divided out the power law $d\ln(\xi/L)/dT \sim L$
of the clean Ising universality class to make the corrections predicted in eq.\ (\ref{eq:dRdT_ln})
more easily visible. The figure demonstrates that these corrections are not of power-law type, instead the data can be fitted well by the predicted logarithmic form
$a [\ln(bL)]^{-1/2}$,  yielding reduced error sums $\bar \chi^2$ of about 0.3 for $\xi_\textrm{av}$ and 0.9 for $\xi_\textrm{gl}$.

\begin{figure}
\includegraphics[width=\columnwidth]{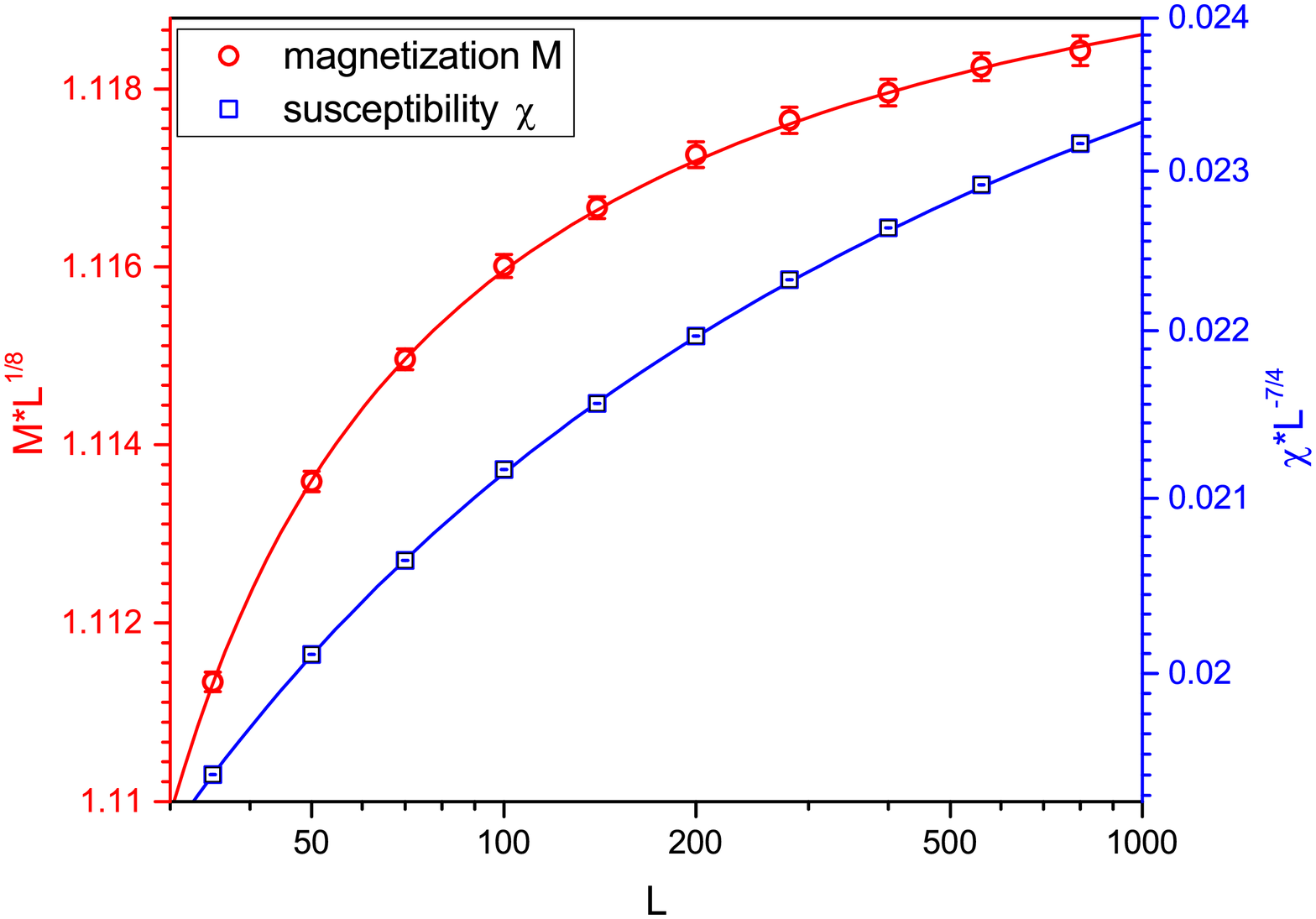}
\caption{Double-logarithmic plots of $M\, L^{1/8}$ and $\chi\,L^{-7/4}$ vs.\ $L$
     at the critical temperature of $T_c=1.9639$. The solid lines are fits to
     $a[1+b/\ln(cL)]$, as suggested by eqs.\ (\ref{eq:M_ln}) and (\ref{eq:chi_ln}).}
\label{fig:Mchi}
\end{figure}

In addition, we study the system size dependence of the magnetization and the magnetic susceptibility at the critical temperature.
According to eqs.\ (\ref{eq:M_ln}) and (\ref{eq:chi_ln}), these quantities are predicted to follow the clean Ising power
laws with additive logarithmic corrections. We again divide out the clean Ising power laws and plot the resulting quantities,
viz., $M\, L^{1/8}$ and $\chi\, L^{-7/4}$ in Fig.\ \ref{fig:Mchi}.
The figure demonstrates that the corrections to the Ising universality class are not of power-law form.
Moreover, they are very weak, in particular for the magnetization where
they change the value by less than 1\% over the entire system size range. To capture this small correction,
we had to increase the number of disorder configurations significantly (500,000 for all system sizes)
to reduce the error bars. The larger numerical effort limits these simulations to system sizes between
$L=35$ and 800. Both  $M\, L^{1/8}$ and $\chi\, L^{-7/4}$ can be fitted well to the function
$a[1+b/\ln(cL)]$ predicted by the renormalization group results (\ref{eq:M_ln}) and (\ref{eq:chi_ln}).
The reduced error sums $\bar \chi^2$ are approximately 0.4 for the susceptibility and 0.3 for the
magnetization.

\section{Summary and conclusions}
\label{sec:conclusions}

In summary, we have carried out large-scale Monte Carlo simulations of the
site-diluted two-dimensional four-color Ashkin-Teller model. Our results confirm that the first-order
phase transition occurring in the undiluted (clean) model is rounded by the disorder,
as required by the Aizenman-Wehr theorem \cite{AizenmanWehr89}.

We have used finite-size scaling of the magnetization, magnetic susceptibility, specific heat
and correlation length to determine the universality class of the emerging continuous phase transition.
All our data agree very well with the results of the perturbative renormalization group
\cite{Murthy87,Cardy96,Cardy99} which predicts critical behavior in the clean two-dimensional
Ising universality class, but with logarithmic corrections similar to those occurring in
the two-dimensional disordered Ising model. These findings agree with those of extensive
high-accuracy simulations of the three-color case \cite{ZWNHV15}. Consequently, they provide strong evidence
for the universality of the critical behavior of the disordered Ashkin-Teller model.

Possible reasons for the discrepancies between our results (Ref.\ \cite{ZWNHV15} and the present paper)
and those of the earlier simulations \cite{BellafardKatzgraberTroyerChakravarty12,BellafardChakravartyTroyerKatzgraber15}
were discussed in detail in Ref.\ \cite{ZWNHV15}. Here, we just reiterate the main points:
Our systems are much larger, up to $2240^2$ sites (compared to just $32^2$ sites
in Ref.\ \cite{BellafardKatzgraberTroyerChakravarty12} and $128^2$ sites in Ref.\
\cite{BellafardChakravartyTroyerKatzgraber15}). This suggests that the earlier simulations
were not in the asymptotic regime, especially for weak disorder where the crossover from
the clean first-order transition to the disordered continuous one is slow. Note, however, that
a discrepancy exists already for the \emph{clean} three-color model. The clean phase diagram
of Ref.\ \cite{ZWNHV15}, which was determined and verified using three independent Monte Carlo algorithms,
coincides with older results by Grest and Widom  \cite{GrestWidom81} but disagrees with
Ref.\ \cite{BellafardKatzgraberTroyerChakravarty12}.

From a more general point of view, the renormalization groups results for the random-$T_c$
Ising \cite{DotsenkoDotsenko83,Shalaev84,Shankar87,Ludwig88} and Ashkin-Teller \cite{Cardy96} models
suggested a kind of ``super-universality'' of critical points in two-dimensional disordered systems.
This idea was initially supported by computer simulations of disordered Ising\cite{ADSW90,WSDA90a}, Ashkin-Teller\cite{WisemanDomany95},
and Potts\cite{WisemanDomany95,ChenFerrenbergLandau95} models as well as interface arguments
\cite{KSSD95}.
Numerical sudies of the disordered $q$-state Potts model\cite{JacobsenCardy98,ChatelainBerche98}
showed, however, that the finite-size scaling exponent $\beta/\nu$ differs from the Ising value and varies with $q$.
Moreover, the phase transition in the random-bond Blume-Capel model was also found to display complex non-Ising behavior, at least for
strong disorder \cite{MBHF09,MBHFP10,TheodorakisPanagiotisFytas12}.

Recently, the disordered quantum Ashkin-Teller spin chain has attracted lots of interest
because it serves as a paradigmatic model for studying disorder effects at first-order
\emph{quantum} phase transitions. As a quantum version of the Aizenman-Wehr theorem
has been proven \cite{GreenblattAizenmanLebowitz09}, the first-order character of the transition
in the clean problem must change upon the introduction of disorder.
Recent strong-disorder renormalization group approaches predict continuous transitions
governed by infinite-randomness critical points in different universality classes, depending on
the coupling strength $\epsilon$ \cite{GoswamiSchwabChakravarty08,HrahshehHoyosVojta12,Barghathietal14}.
Furthermore, in the case of two colors an exotic strong-disorder infinite-coupling phase \cite{HHNV14}
is predicted to appear for large $\epsilon$. These predictions can be tested by generalizations
of our Monte Carlo method to the  $(1+1)$-dimensional quantum case.
Some work along these lines is already in progress.

It may also be interesting to investigate the Ashkin-Teller model defined on a topologically
disordered lattice such as the random Voronoi-Delaunay lattice (see, e.g., \cite{OBSC_book00}).
Recent work \cite{BarghathiVojta14} has shown that the Imry-Ma argument does not hold for these
lattices because a topological constraint suppresses the disorder fluctuations. This leaves
open the possibility that the first-order phase transition survives in the presence of such
topological disorder.

This work was supported in part by the NSF under Grant Nos.\ DMR-1205803 and DMR-1506152.


\end{document}